\documentclass[preprint]{elsarticle}

\usepackage{epsfig}
\usepackage{amsmath}
\begin{document}

\newcommand{\bvec}[1]{\mbox{\boldmath ${#1}$}}

\title{Effects of the higher partial waves and relativistic terms
 on the accuracy of the calculation of the 
 hypertriton electroproduction}

\author{T. Mart}

\address{Departemen Fisika, FMIPA, Universitas Indonesia, Depok 16424, Indonesia}

\begin{abstract}
We have investigated the accuracies of calculations made by omitting 
the higher partial waves of nuclear wave functions and the elementary 
relativistic terms in the hypertriton
electroproduction. We found that an accurate calculation 
would still be obtained if we used at least three lowest partial waves
with isospin $T=0$. Furthermore,
we found that the omission of the relativistic terms in the elementary
process amplitude could lead to a large deviation from the full calculation.
We also present the cpu-times required to calculate the cross sections.
For future consideration 
the use of these lowest partial waves is suggested, since the calculated
cross section deviates only about 0.17 nb/sr ($\approx 4$\%), at most, from the 
full calculation, whereas the cpu-time is reduced by a factor of 60.
Comparison of our result with the available experimental data supports these findings.
\end{abstract}


\maketitle

\section{Introduction}
An accurate calculation using a simple formalism is naturally desired
in all phenomenological studies of nuclear and particle physics. However,
in most cases this is difficult to achieve, because an accurate calculation
usually does not allow for any extreme approximation. 
Therefore, an optimal approximation which only includes the most
important parts of the formalism, without sacrificing the accuracy
of the numerical result, should be obtained. 
A good example is found in 
the analysis of the hypertriton photo- and electroproduction, i.e.,
\begin{eqnarray*}
  \gamma+ {^3{\rm He}}\to {K^+}+ {^3_\Lambda{\rm H}} ~~~{\rm and}~~~
  e+ {^3{\rm He}}\to e'+{K^+}+ {^3_\Lambda{\rm H}} ~,
\end{eqnarray*}
given in Refs.~\cite{Mart:2008gq,mart98}. In this analysis the cross section
is calculated by means of the elementary operator 
Kaon-Maid~\cite{kaon-maid} sandwiched
between two nuclear wave functions ($^3$He and ${^3_\Lambda}$H)
obtained from the solutions of Faddeev equations using modern
nucleon-nucleon and hyperon-nucleon potentials~\cite{nijmegen93,miyagawa93}.
The numbers of partial waves for the $^3$He and ${^3_\Lambda}$H 
wave functions are 34 and 16, respectively. Along with the corresponding
probabilities these partial waves are shown in Table~\ref{probability_wf}.
In both wave functions the total numbers of supporting points for the 
two-body ($\bvec{p}$) and the spectator ($\bvec{q}$) momenta are 34 
and 20, respectively. As we shall see in the next Section, this could
lead to a problem of integration with almost two billions grid points.
Therefore, it is obviously very important to limit the number of participating 
partial waves or to truncate the elementary amplitude in order to 
simplify the formalism as well as to avoid the unnecessarily long cpu-time 
required to calculate the reaction cross section. 
At $W=4.04$ GeV  it has  been shown in Ref.~\cite{Mart:2008gq} that for the 
hypertriton photoproduction the use of four lowest partial waves 
($\alpha\le 4$, see Table~\ref{probability_wf} for the explanation 
of $\alpha$) would nicely approximate
the full calculation, whereas the use of $\alpha\le 5$ would lead to
a perfect result. It has been also pointed out that
careful inspections in a wide range of kinematics should be 
performed, before we can apply this approximation 
in the hypertriton photo- and electroproduction~\cite{Mart:2008gq}. 

Furthermore, it has
been also known that the hypernucleus production cross section is sensitive
to the elementary amplitude, especially at the forward directions,
where the two recent experimental data sets from SAPHIR  \cite{SP03} 
and CLAS \cite{CL05} collaborations show a lack of mutual 
consistency \cite{Bydzovsky:2006wy}. As a consequence, 
hypertriton production at this kinematics could also shed light on 
the solution of this discrepancy problem. However, the extraction
of the information on the elementary amplitude from the nuclear
cross sections requires a massive fitting process, which would
become impossible if the cpu-times required to calculate these cross 
sections were extremely long.

The present analysis is greatly motivated by these facts. 
Here we shall quantitatively investigate the effects of 
omitting the higher partial waves of nuclear wave functions on 
the accuracy of the calculation. We shall also compare this result
with the result of the approximation made by excluding the
relativistic terms in the elementary amplitude as suggested
in Ref.~\cite{Adelseck:1986fb}. To this end we take the 
electroproduction process, since photoproduction is only 
a special case of electroproduction. 
Our motivation is obvious, namely to find the shortest cpu-time for which
the deviation of the calculated cross section from the full calculation
is still controllable. 

This paper is organized as follows: In Section 
\ref{sec:formalism} we shall briefly review our formalism and state
our problem. We shall present and discuss the results of our calculations 
in Section \ref{sec:result}. Section \ref{sec:compare_data} will focus
on the comparison of our results with experimental data. In 
Section \ref{sec:conclusion} we shall summarize our findings. 

\begin{table}[!ht]
\renewcommand{\arraystretch}{1.0}
\begin{center}
\caption{Quantum numbers and probabilities (in \%) of the $^3$He and 
the hypertriton wave functions \protect\cite{nijmegen93,miyagawa93}. }
\label{probability_wf}
\begin{tabular}{cccccccrr}
\hline\hline
\\[-2ex]
$~~\alpha~~$ &$~~L~~$ & $~~S~~$ & $~~J~~$ & $~~l~~$ & $~~2j~~$ & $~~2T~~$ &
~~$P(^3{\rm He})$~~&~~$P(^3_{\Lambda}{\rm H})$~~ \\
[1ex]
\hline\\[-2ex]
 1 & 0 & 0 & 0 & 0 & 1 & 1 & 44.580  &   -~~~   \\
 2 & 0 & 1 & 1 & 0 & 1 & 0 & 44.899  &  93.491 \\
 3 & 2 & 1 & 1 & 0 & 1 & 0 &  2.848  &   5.794 \\
 4 & 0 & 1 & 1 & 2 & 3 & 0 &  0.960  &   0.034 \\
 5 & 2 & 1 & 1 & 2 & 3 & 0 &  0.189  &   0.027 \\
 6 & 1 & 0 & 1 & 1 & 1 & 0 &  0.089  &   0.004 \\
 7 & 1 & 0 & 1 & 1 & 3 & 0 &  0.198  &   0.008 \\
 8 & 1 & 1 & 0 & 1 & 1 & 1 &  1.107  &   -~~~   \\
 9 & 1 & 1 & 1 & 1 & 1 & 1 &  1.113  &   -~~~   \\
10 & 1 & 1 & 1 & 1 & 3 & 1 &  0.439  &   -~~~   \\
11 & 1 & 1 & 2 & 1 & 3 & 1 &  0.064  &   -~~~   \\
12 & 3 & 1 & 2 & 1 & 3 & 1 &  0.306  &   -~~~   \\
13 & 1 & 1 & 2 & 3 & 5 & 1 &  1.018  &   -~~~   \\
14 & 3 & 1 & 2 & 3 & 5 & 1 &  0.024  &   -~~~   \\
15 & 2 & 0 & 2 & 2 & 3 & 1 &  0.274  &   -~~~   \\
16 & 2 & 0 & 2 & 2 & 5 & 1 &  0.425  &   -~~~   \\
17 & 2 & 1 & 2 & 2 & 3 & 0 &  0.122  &   0.024 \\
18 & 2 & 1 & 2 & 2 & 5 & 0 &  0.095  &   0.018 \\
19 & 2 & 1 & 3 & 2 & 5 & 0 &  0.205  &   0.053 \\
20 & 4 & 1 & 3 & 2 & 5 & 0 &  0.053  &   0.006 \\
21 & 2 & 1 & 3 & 4 & 7 & 0 &  0.126  &   0.010 \\
22 & 4 & 1 & 3 & 4 & 7 & 0 &  0.038  &   0.007 \\
23 & 3 & 0 & 3 & 3 & 5 & 0 &  0.005  &   0.001 \\
24 & 3 & 0 & 3 & 3 & 7 & 0 &  0.008  &   0.001 \\
25 & 3 & 1 & 3 & 3 & 5 & 1 &  0.051  &   -~~~   \\
26 & 3 & 1 & 3 & 3 & 7 & 1 &  0.045  &   -~~~   \\
27 & 3 & 1 & 4 & 3 & 7 & 1 &  0.008  &   -~~~   \\
28 & 5 & 1 & 4 & 3 & 7 & 1 &  0.074  &   -~~~   \\
29 & 3 & 1 & 4 & 5 & 9 & 1 &  0.178  &   -~~~   \\
30 & 5 & 1 & 4 & 5 & 9 & 1 &  0.006  &   -~~~   \\
31 & 4 & 0 & 4 & 4 & 7 & 1 &  0.053  &   -~~~   \\
32 & 4 & 0 & 4 & 4 & 9 & 1 &  0.059  &   -~~~   \\
33 & 4 & 1 & 4 & 4 & 7 & 0 &  0.011  &   0.004 \\
34 & 4 & 1 & 4 & 4 & 9 & 0 &  0.009  &   0.003 \\
[1ex]
\hline\hline
\end{tabular}
\end{center}
\end{table}

\section{Formalism and Statement of the Problem}
\label{sec:formalism}
\begin{figure}[!t]
  \begin{center}
    \leavevmode
    \epsfig{figure=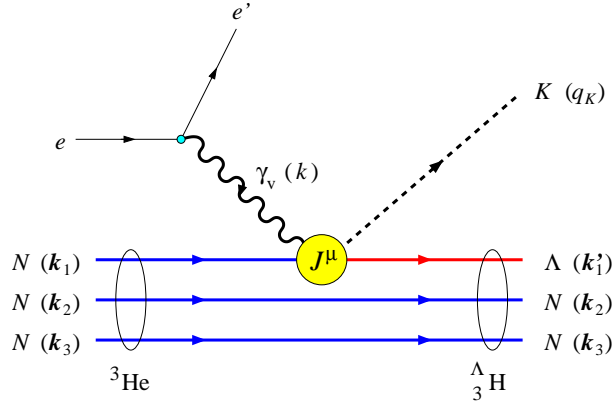,width=80mm}
    \caption{Electroproduction of the hypertriton
        on a $^3$He target 
        in an impulse approximation, where the virtual photon
        interacts with only one nucleon inside the $^3$He. }
   \label{fig:impulse} 
  \end{center}
\end{figure}

The formalism of the hypertriton electroproduction
off $^3$He in an impulse approximation has been presented in 
Ref.~\cite{Mart:2008gq}. Except for the difference between
the initial and final nuclear masses, as well as between the
initial proton and the final hyperon masses in the elementary
operator, the formulas are similar to those used in pion
electroproduction off $^3$He \cite{tiator81}. 
To facilitate the discussion, here we will 
only present the most important part of them.
We start with the corresponding nuclear transition matrix 
element, which can be written 
as~\cite{Mart:2008gq}
\begin{eqnarray}
  \langle\, {\rm {^3_\Lambda}H} \left| \, J^\mu \, \right| {\rm {^3{\rm He}}} 
  \,\rangle =
  \sqrt{3}\,\int d^{3}\bvec{p}~d^{3}\bvec{q}~ \Psi_{\rm {^3_\Lambda}H}^{*}(\bvec{p}, 
  \bvec{q}') ~ J^\mu\, (\bvec{k},\bvec{k}_{1}, \bvec{k}_1') ~
  \Psi_{\rm {^3{\rm He}}}(\bvec{p},\bvec{q})~ ,~ 
  \label{trans1}
\end{eqnarray}
where the factor of
$\sqrt{3}$ on the right hand side of Eq.\,(\ref{trans1}) comes from the 
anti-symmetry of the initial state, $J^\mu$ represents the elementary
operator, while  
the integrations are taken over the three-body momentum coordinates
(see Fig.~\ref{fig:impulse} for the explanation of the momenta)
\begin{eqnarray}
  \bvec{p} ~=~ \frac{1}{2}\,(\bvec{k}_2 - \bvec{k}_3) ~~ , ~~
  \bvec{q} ~=~ \bvec{k}_1 ~ ,
\end{eqnarray}
and the hyperon momentum in the hypertriton is given by
\begin{eqnarray}
  \bvec{q}' &=& \bvec{k}_1 + \frac{2}{3}\,(\bvec{k}-\bvec{q}_K)~.
\end{eqnarray}

By expanding the three-body wave functions in Eq.~(\ref{trans1}) in 
terms of their orbital momenta, spins, and isospins, 
\begin{eqnarray}
\Psi(\bvec{p},\bvec{q}) & = & \sum_{\alpha,{\sf m}}
\phi_{\alpha} (p,q) (Lm_{L}Sm_{S}|J m_{J})
(l m_{l}{\textstyle \frac{1}{2}} m_{s}|j m_{j})\,(J m_{J} j m_{j}
|{\textstyle \frac{1}{2}} M_{\rm i})  \nonumber\\
&& \times 
Y^{L}_{m_{L}} (\bvec{\hat{p}})Y^{l}_{m_{l}}(\bvec{\hat{q}})
\chi^{S}_{m_{S}}\chi_{m_{s}}^{\frac{1}{2}}~ \left|\,
(T{\textstyle \frac{1}{2}}){\textstyle \frac{1}{2}} M_{t}~\right\rangle ~ ,
\label{wfhe3}
\end{eqnarray}
we can recast the transition matrix element in Eq.~(\ref{trans1}) into the form
\begin{eqnarray}
  \langle\, {\rm {^3_\Lambda}H} \left| \, J^\mu \, \right| {\rm {^3{\rm He}}} \,\rangle &=&
  \sqrt{3}\, \sum_{\alpha , \alpha '}
  \sum_{{\sf mm}'}
        \left(Lm_{L}Sm_{S}|Jm_{J}\right) \, \nonumber\\
&& \times\,\left(Lm_{L}Sm_{S}|J'm_{J'}\right) \,
\left(lm_{l}{\textstyle \frac{1}{2}}m_{s}|jm_{j}\right) \nonumber\\
&& \times\,\left(l'm_{l'}{\textstyle \frac{1}{2}}m_{s'}|j'm_{j'}\right) \, 
        \left(Jm_{J}jm_{j}|{\textstyle \frac{1}{2}}M_{\rm i}\right) \, 
        \nonumber\\
&& \times\,\left(J'm_{J'}j'm_{j'}|{\textstyle \frac{1}{2}}M_{\rm f}\right) 
\delta_{LL'}\,\delta_{m_{L}m_{L'}}\,
           \delta_{SS'}\,\delta_{m_{S}m_{S'}}\,\delta_{T0}
   \nonumber\\
&& \times\,\int p^{2}dp~d^{3}\bvec{q}\;\phi_{\alpha '}(p,q')\;
   \phi_{\alpha}(p,q)\; Y^{l'}_{m_{l'}}(\bvec{\hat q}')
   \, Y^{l}_{m_{l}}(\bvec{\hat q})\;   \nonumber\\
&& \times\, \langle {\textstyle \frac{1}{2}},m_{s'}\, 
   |\, J^{\mu} \,|\, {\textstyle \frac{1}{2}},m_{s} \rangle \, ,
\label{trans2}
\end{eqnarray}
where $\alpha$ and $\alpha'$ indicate the partial waves of the
initial and final nuclear wave functions, respectively, while the notations 
${\sf m} = (m_{L}m_{S}m_{l}m_{s}m_{J}m_{j})$ and
${\sf m}' = (m_{L'}m_{S'}m_{l'}m_{s'}m_{J'}m_{j'})$ 
have been introduced for the sake of brevity. 

From Table~\ref{probability_wf} we can estimate that a full calculation
of the four-dimensional integrals in Eq.~(\ref{trans2}) 
using all partial waves could involve integrations with 
$34\times 16\times 34\times 20\times 30\times 10=110,976,000$ grid 
points, where the two last numbers in the multiplications 
($30\times 10$) come from the minimum Gauss supporting 
points for the numerically-stable angular integrations~\cite{Mart:2008gq}.

It is also important to note that these integrations are performed over
all components of 
the transition matrix element in the form of $4\times 4$ matrix
$[\,j^{\mu}\,]^{(n)}_{m_n}$ (or equivalently $[\,j_{\mu\nu}\,]$), where 
\begin{eqnarray}
  \label{eq:jmu_def}
  J^\mu &=& \sum_{n=0}^1\;\sum_{m_n=-n}^{+n}\, 
            (-1)^{m_n}\,\sigma^{(n)}_{-m_n}\,
            [\,j^{\mu}\,]^{(n)}_{m_n} \nonumber\\
          &=& (1,\sigma_x,\sigma_y,\sigma_z) 
        \left(\begin{array}{cccc}
        j_{00} & j_{x0} & j_{y0} & j_{z0}\\
        j_{0x} & j_{xx} & j_{yx} & j_{zx}\\
        j_{0y} & j_{xy} & j_{yy} & j_{zy}\\
        j_{0z} & j_{xz} & j_{yz} & j_{zz}
        \end{array} \right)~.
\end{eqnarray}
As a consequence, the problem of calculating the cross section 
becomes numerically more challenging, since it is equivalent to
the problem of integration with 1,775,616,000 grid points.
Furthermore, the result of this
integration must be summed over angular-momentum and spin 
projections $m_{J},m_{J'},m_S$, and $m_s$ [indicated by 
{\sf m} and ${\sf m}'$ in  Eq.~(\ref{trans2})]. 
Fortunately, the selection rule represented by the three Kronecker
delta functions in Eq.~(\ref{trans2})
along with current conservation 
reduce this number to about 156 millions grid points. Nevertheless,
this still indicates a time-consuming numerical computation.
 
On the other hand, the elementary operator of the 
 elementary process $\gamma_v(k)+p(p_p)\to {K^+}(q_K)+\Lambda(p_\Lambda)$
can be written as 
\begin{eqnarray}
\label{nro1}
\lefteqn{ \langle\, \Lambda\,| \,\epsilon_\mu\,J^\mu\,|\, p \,\rangle
 = \sqrt{\frac{\varepsilon_p\varepsilon_\Lambda}{4m_pm_\Lambda}}\, 
\chi_{\Lambda}^{\dagger}\, \biggl[\, F_{1}
\,{\bvec{\sigma}} \cdot {\bvec{\epsilon}}
+ F_{2}\,  {\bvec{\sigma}} \cdot {\bvec k}\, \epsilon_{0}
+ F_{3}\, {\bvec{\sigma}} \cdot {\bvec k}\, 
{\bvec k} \cdot  {\bvec{\epsilon}}  }
 \nonumber\\
 & &+ F_{4}\, 
{\bvec{\sigma}} \cdot {\bvec k}\, {\bvec p}_{p} \cdot 
{\bvec{\epsilon}}+ F_{5}\, {\bvec{\sigma}}
\cdot {\bvec k}\, {\bvec p}_{\Lambda} \cdot {\bvec{\epsilon}}
+ \frac{1}{\varepsilon_p}\Bigl\{
F_{6}\, {\bvec{\sigma}}\cdot {\bvec p}_{p}\,\epsilon_{0} 
+ F_{7}\, {\bvec{\sigma}} 
\cdot {\bvec p}_{p}\, {\bvec k} \cdot {\bvec{\epsilon}}
\nonumber\\
 & &
+ F_{8}\, {\bvec{\sigma}} \cdot {\bvec p}_{p}\, {\bvec p}_{p} 
\cdot {\bvec{\epsilon}} 
+ F_{9}\, 
{\bvec{\sigma}} \cdot {\bvec p}_{p}\, 
{\bvec p}_{\Lambda} \cdot  {\bvec{\epsilon}}
+ F_{14}\, {\bvec{\sigma}} \cdot 
{\bvec{\epsilon}}\, {\bvec{\sigma}}
 \cdot {\bvec k}\, {\bvec{\sigma}} \cdot {\bvec p}_{p}
\Bigr\} 
\nonumber\\ 
&& + \frac{1}{\varepsilon_\Lambda}\Bigl\{ F_{10}
 {\bvec{\sigma}} \cdot {\bvec p}_{\Lambda}\, 
\epsilon_{0} 
+ F_{11}\, {\bvec{\sigma}} \cdot {\bvec p}_{\Lambda}\, {\bvec k} \cdot 
{\bvec{\epsilon}} + F_{12}\, {\bvec{\sigma}}
 \cdot {\bvec p}_{\Lambda}\,
 {\bvec p}_{p} \cdot {\bvec{\epsilon}}
\nonumber\\  & &   
+ F_{13} {\bvec{\sigma}} \cdot {\bvec p}_{\Lambda}\,
{\bvec p}_{\Lambda} \cdot  {\bvec{\epsilon}}
+ F_{15}\, {\bvec{\sigma}} \cdot {\bvec p}_{\Lambda}\, 
{\bvec{\sigma}} \cdot {\bvec{\epsilon}}\, {\bvec{\sigma}} \cdot {\bvec k}
\Bigr\}
\nonumber\\  & &
+ \frac{1}{\varepsilon_p\,\varepsilon_\Lambda}\Bigl\{F_{16}
 {\bvec{\sigma}} \cdot {\bvec p}_{\Lambda}\, 
{\bvec{\sigma}} \cdot {\bvec{\epsilon}}\, {\bvec{\sigma}} \cdot {\bvec p}_{p} 
+\, F_{17}\, {\bvec{\sigma}} \cdot {\bvec p}_{\Lambda}\, 
 {\bvec{\sigma}} \cdot 
{\bvec k}\,{\bvec{\sigma}} \cdot {\bvec p}_{p}\, \epsilon_{0} 
\nonumber\\&&
+ F_{18} {\bvec{\sigma}} \cdot {\bvec p}_{\Lambda}\, 
{\bvec{\sigma}} \cdot {\bvec k}\, {\bvec{\sigma}} 
\cdot {\bvec p}_{p}\, {\bvec k} \cdot {\bvec{\epsilon}}
+ F_{19}\, {\bvec{\sigma}} \cdot {\bvec p}_{\Lambda}\,
{\bvec{\sigma}} \cdot {\bvec k}\, {\bvec{\sigma}} 
\cdot {\bvec p}_{p}\, {\bvec p}_{p} \cdot  {\bvec{\epsilon}}
\nonumber\\ 
 & & 
+\, F_{20}\, {\bvec{\sigma}} \cdot {\bvec p}_{\Lambda}\, 
{\bvec{\sigma}} \cdot {\bvec k}\, {\bvec{\sigma}} 
\cdot {\bvec p}_{p}\, {\bvec p}_{\Lambda} \cdot  {\bvec{\epsilon}}\Bigr\}
 \biggr] \chi_{p} ~,
\end{eqnarray}
where the individual amplitudes $F_{i}$ are given in 
Refs.\,\cite{Mart:2008gq,mart98}, $\epsilon_\mu$ is the virtual photon
polarization vector, $\varepsilon_p=E_{p} + m_{p}$, and 
$\varepsilon_\Lambda=E_{\Lambda} + m_{\Lambda}$.

From Eq.~(\ref{nro1}) it is obvious that the amplitudes $F_{16}-F_{20}$ 
and $F_{6}-F_{15}$  originate from the ``small-small'' (SS)
and  ``small-big'' (SB) terms of the Dirac spinors, respectively, 
whereas the rest correspond to the ``big-big'' (BB) terms.
It can be easily shown that the ratio between the ``small'' 
and ``big'' terms is of the order $v/c$ (see, e.g., Ref.~\cite{halzen}). 
In pion
photoproduction near threshold it has been widely known that only
the leading Kroll-Ruderman term $F_1$ significantly contributes to
the process. 
However, in kaon photo- and electroproduction the high
threshold energy of the process could certainly change this picture. 
In view of this, it is certainly tempting to neglect the relativistic 
SS-terms in our formalism as well as to investigate the effects of the  
SB- and other terms on the calculated cross sections of the hypertriton 
electroproduction. 

For the purpose of numerical computation we note that 
the cross section of the hypertriton electroproduction 
in the c.m. system can be written as
\begin{eqnarray}
  \label{eq:ds_domega_virtual}
  \frac{d\sigma_v}{d\Omega_K} = \frac{d\sigma_{\rm T}}{d\Omega_K} +
  \epsilon_{\rm L}~ \frac{d\sigma_{\rm L}}{d\Omega_K} +
  \epsilon~ \frac{d\sigma_{\rm TT}}{d\Omega_K}~ \cos 2\phi_K +
  \sqrt{2\epsilon_{\rm L}(1+\epsilon)}~ 
  \frac{d\sigma_{\rm LT}}{d\Omega_K}~ \cos \phi_K ~,
\end{eqnarray}
with $\epsilon$ represents the virtual photon polarization,
$\epsilon_L=-(k^2/\bvec{k}^2)\,\epsilon$  and
the individual cross sections can be written as
\begin{eqnarray}
  \frac{d\sigma_i}{d\Omega_K} = \alpha_e \,
  \frac{q_{K}}{K_L}\, \frac{M_{^3_\Lambda\!{\rm H}}}{2W}\, 
  W_i ~~ ,~~(i={\rm T},{\rm L},{\rm TT}, {\rm and~LT})~,
\end{eqnarray}
with $\alpha_e=e^2/4\pi$ and $K_L = (W^2-M_{\rm He}^2)/2M_{\rm He}$.
The nuclear structure functions $W_i$ are given by 
\begin{eqnarray}
  W_{\rm T} &=& \frac{1}{4\pi}~ (W^{xx}+W^{yy}) ~,\\
  W_{\rm L} &=& \frac{1}{4\pi}~ W^{00} ~,\\
  W_{\rm TT} &=& \frac{1}{4\pi}~ (W^{xx}-W^{yy}) ~,\\
  W_{\rm LT} &=& \frac{1}{4\pi}~ (W^{0x}+W^{x0}) ~,
\end{eqnarray}
where the spin averaged Lorentz tensor $W^{\mu\nu}$
is related to the nuclear transition matrix 
element given in Eq.~(\ref{trans1}) by
\begin{eqnarray}
  W^{\mu\nu} &=& \frac{1}{2} \sum_{s_{\rm i}s_{\rm f}}
  \langle\, {\rm {^3_\Lambda}H} \left| \, J^\mu \, \right| {\rm {^3{\rm He}}} 
  \,\rangle\,
  \langle\, {\rm {^3_\Lambda}H} \left| \, J^\nu \, \right| {\rm {^3{\rm He}}} 
  \,\rangle ^* ~.
\end{eqnarray}

The kinematic chosen in this analysis is close
to that of experimental data~\cite{Dohrmann:2004xy}, 
because we want to explore the kinematics region that is 
experimentally accessible. The angular distribution will be
limited in the range of 
$0^\circ\le\theta_K\le 31^\circ$, in which the magnitude of the 
cross section is sizeable. The total c.m. energy is limited to
3.5 GeV $\le W\le$ 5 GeV, since below this limit the cross section
would be too small, whereas above this limit the elementary operator
would lose its predictive power.

\section{Results and Discussion}
\label{sec:result}
\begin{figure}[!t]
    \epsfig{figure=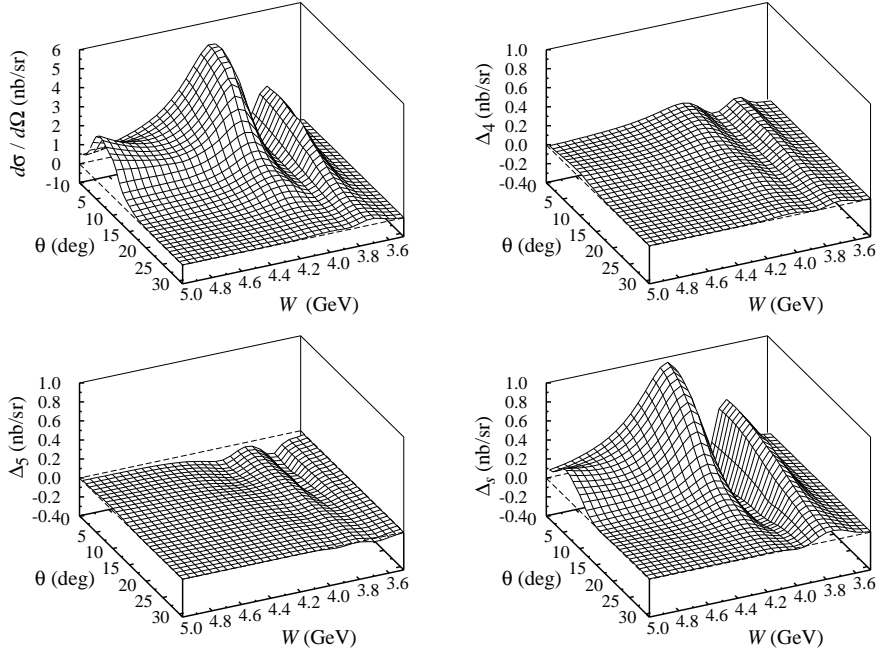,width=120mm}
    \caption{Effects of the higher partial waves on the differential cross
        section of the hypertriton electroproduction off $^3$He
        at $k^2=-0.35$ GeV$^2$ and $\epsilon=0.762$. The upper-left panel
        shows the result obtained from the full calculation 
        by using all partial waves. Other panels
        display the differences between the full calculation and the calculation
        by using $\alpha\le 5$ ($\Delta_5$), $\alpha\le 4$ ($\Delta_4$), and  
        only $s$-waves ($\Delta_s$). Note that for the sake of visibility 
        a different scale has been used for the vertical axis of the 
        upper-left panel.}
   \label{fig:pwaves} 
\end{figure}

Figure \ref{fig:pwaves} demonstrates the cross section obtained from
the full calculation using all partial waves and the deviations from
this result if we use $\alpha\le 5$, $\alpha\le 4$, and $s$-waves 
($\alpha=2,4$). Our experience shows that computing the numerical 
data required by the plot of the differential cross section shown 
in the upper-left panel of Fig.~\ref{fig:pwaves} (consisting of $31\times 31=961$ points) on 
a PC with a single processor Pentium-4 takes about 11 days (15,344 min). 

From the lower-left panel of Fig.~\ref{fig:pwaves} it is obvious that
limiting the partial waves up to $\alpha=5$ yields an accurate
approximation, since in general it just slightly underestimates the full
calculation. The largest discrepancies are found at the two cross
section peaks at $W\approx 3.75$ GeV and 4.10 GeV close to the forward
angle, i.e., about $0.15$ nb/sr (less than 3\%). The average deviation over these 961
points is only 0.019 nb/sr. 
A similar behavior is also found if we use $\alpha\le 4$,
except in this case the calculated cross section 
slightly overestimates the cross section of
the full calculation. Here, the average and largest deviations 
are found to be 0.038 nb/sr and 0.17 nb/sr (less than 4\%), respectively.
Finally, the largest deviation, almost 1 nb/sr at the top of the 
highest cross section peak, is obtained if we use only $s$-waves. 
Since the largest
differential cross section is around 5 nb/sr, it is obvious that the
latter provides a relatively poor approximation method for the hypertriton 
electroproduction.

\begin{figure}[!t]
  \begin{center}
    \leavevmode
    \epsfig{figure=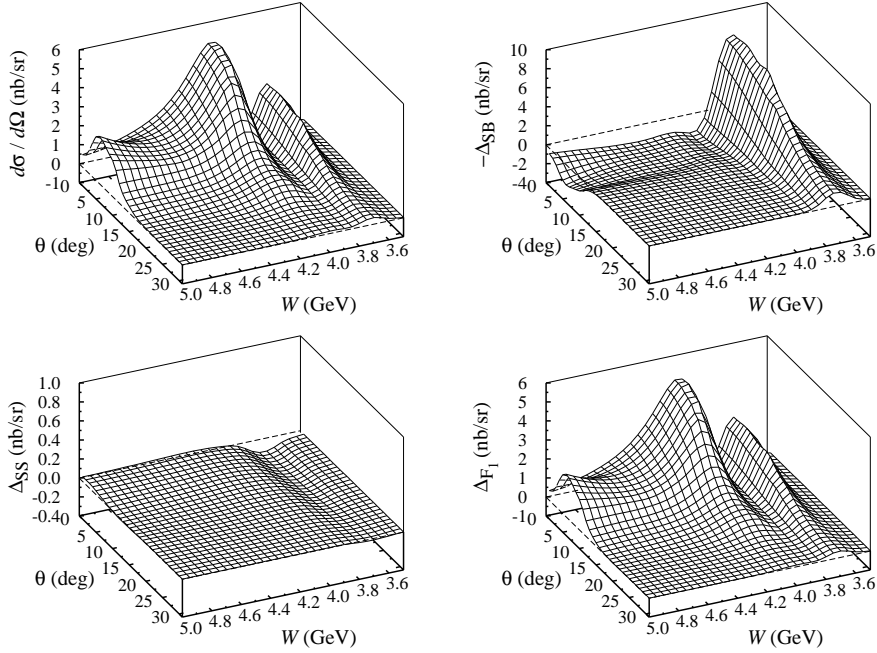,width=120mm}
    \caption{Effects of the relativistic terms on the differential cross
        section of the hypertriton electroproduction off $^3$He. The upper-left panel
        shows the result of a full calculation 
        obtained by using all terms in Eq.~(\ref{nro1}), i.e.,
        $F_1$-$F_{20}$. Other panels display the differences between the full 
        calculation and the calculation 
        by using $F_1$-$F_{15}$ ($\Delta_{\rm SS}$), 
        $F_1$-$F_{5}$ ($\Delta_{\rm SB}$), and only $F_1$ ($\Delta_{F_1}$). 
        Note that all figures are obtained by using $\alpha\le 5$ and 
        the same kinematics as in Fig.~\ref{fig:pwaves}.
        For the sake of visibility different scales have been used
        in vertical axes.}
   \label{fig:relativ} 
  \end{center}
\end{figure}

Figure~\ref{fig:relativ} demonstrates the effects of 
excluding the relativistic terms on the calculated 
differential cross section. Since it has just been shown
that the use of $\alpha\le 5$ can approximate the
full calculation with an accuracy up to 3\%, we believe 
that it is sufficient to limit the use of partial waves up 
to $\alpha\le 5$ at this stage. 

The lower-left panel of 
Fig.~\ref{fig:relativ} demonstrates the effect of the
SS-terms exclusion. We note that the maximum deviation
is only about 4\% at $W\approx 3.75$ GeV. Therefore, 
the SS-terms can be safely neglected for this process. 

The shape of the differential cross section obtained 
by excluding the SB-terms is found to be completely different. 
As a consequence, the deviation from the full 
calculation is quite large (see also Table~\ref{tab:cpu_time}).
The extreme nonrelativistic approximation ($F_1$ only)
also yields a large deviation from the full calculation.
The reason that the deviation in the latter is relatively 
smaller than in the former is that this  
approximation yields very small cross section. 
Although the analysis of Ref.~\cite{Adelseck:1986fb} has been
performed at the elementary level and using a quite different
elementary operator, our result corroborates 
its finding, i.e., both the exclusion
of the SB-terms and the extreme nonrelativistic approach 
can not be used as a good approximation in the hypertriton
electroproduction. Furthermore since the exclusion of the
SB-terms was not investigated by the authors of Ref.~\cite{Adelseck:1986fb},
our analysis therefore provides an extension of their finding.

At this stage, it is also important to consider the cpu-times required
to make the plots just shown. Along with the corresponding 
deviations from the full calculation,
the required cpu-times to calculate the 961 points of the
cross section 
by using a single processor Pentium-4 PC 
are listed in Table~\ref{tab:cpu_time}. 

It is interesting to see that
the cpu-time is significantly reduced by a factor of 30 if we
limit the partial waves up to $\alpha=5$, while the accuracy is still
maintained up to about $0.15$ nb/sr. As a consequence, to obtain the
plot shown at the top panel of Fig.~\ref{fig:relativ} we need less
than 9 hours. If we used the partial waves with $\alpha\le 4$,
the cpu-time is reduced by a factor of about 60, whereas
the maximum deviation slightly increases 
to $0.17$ nb/sr. The use of only $s$-waves substantially reduces
the cpu-time, i.e., by a factor of 300. 
The
average deviations displayed in Table~\ref{tab:cpu_time} 
show in general the same behavior.

\begin{table}[!t]
\renewcommand{\arraystretch}{1.3}
\begin{center}
\caption{The cpu-time ($\tau$) required to compute the 
        numerical data 
        of the cross section plot shown at the top panel of 
        Fig.~\ref{fig:pwaves} along with the average and maximum 
        deviations from the full calculation (indicated by $\Delta_{\rm av.}$
        and $\Delta_{\rm max.}$, respectively) for different approximations, 
        i.e., using all partial waves (Full), $\alpha\le 5$ (1), 
        $\alpha\le 4$ (2), and only $s$-waves (3). With $\alpha\le 5$
        the role of different elementary amplitudes $F_i$ given in 
        Eq.~(\protect\ref{nro1}) is demonstrated, i.e., using only 
        $F_1-F_{15}$ (a), $F_1-F_{5}$ (b), and $F_1$ (c).
\label{tab:cpu_time}}
\begin{tabular}{lccccccc}
\hline\hline
 & ~~Full~~~ & ~~~1~~~ & ~~~2~~~ & ~~~3~~~ & ~~~a~~~ & ~~~b~~~ & ~~~c\\
[1ex]
\hline
$\tau$ (min) & 15,344 & 511 & 294 & 46 & 498 & 487 & 485 \\
$\Delta_{\rm av.}$ (nb/sr) & - & 0.02 & 0.04 & 0.14 & 0.02 & 0.98 & 0.81 \\
$\Delta_{\rm max.}$ (nb/sr)~ & - & 0.15 & 0.17 & 0.99 
                   & 0.10 & 8.03 & 4.97\\
\hline\hline
\end{tabular}
\end{center}
\end{table}

In contrast to the omission of the higher partial waves, the exclusion
of the relativistic terms in Eq.~(\ref{nro1}) does not significantly
reduce the cpu-times. This is clearly demonstrated in the first line
of Table~\ref{tab:cpu_time}, where the cpu-time for the calculation
with $\alpha\le 5$ decreases from 511~min (full terms) to 485~min (only $F_1$).
This result proves that computing the rest 19 $F_i$ amplitudes 
in Eq.~(\ref{nro1}) is much simpler, and therefore much faster, 
than computing the massive integrals given in Eq.~(\ref{trans2})
with the complete partial waves.
At the same time the maximum deviation in the latter becomes almost 5 nb/sr.
This is because in the electromagnetic production of kaon the high energy 
of the process enhances the role of all but the  Kroll-Ruderman term
[see  Eq.~(\ref{nro1})], and the extreme  nonrelativistic approach
results in a very small cross section. Consequently, the largest 
difference with the full calculation is about 5 nb/sr, i.e., 
the largest differential cross section found with the full calculation.
Note that the decrease of $\Delta_{\rm max.}$ from 0.15 nb/sr (full terms) 
to 0.10 nb/sr (using only $F_1-F_{15}$) seems to be fortuitous, because 
the average deviations for the two cases are the same (0.02 nb/sr). 

To complete our analysis we have also investigated the effects of the higher
partial waves and the relativistic terms at $k^2=-1.0$ GeV$^2$. However, 
since the obtained cross sections are quite small (0.32 nb/sr, at most),
it is very hard to draw a quantitative conclusion at this kinematics.

\section{Comparison with experimental data}
\label{sec:compare_data}
\begin{figure}[!t]
  \begin{center}
    \leavevmode
    \epsfig{figure=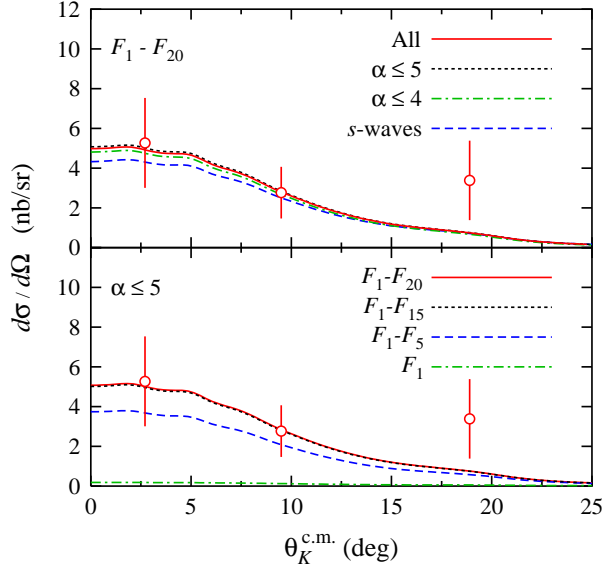,width=80mm}
    \caption{(Color online) Comparison between experimental data
      \protect\cite{Dohrmann:2004xy}
      and the calculation using all and specific numbers of partial
      waves (upper panel) and the approximations made by 
      including only certain elementary amplitudes
      in the elementary operator (lower panel).}
   \label{fig:experiment} 
  \end{center}
\end{figure}

Although our primary motivation is to quantitatively study the
effects of excluding higher partial waves and relativistic terms,
it is also imperative to compare the results with the available 
experimental data given in Ref.~\cite{Dohrmann:2004xy}. The comparison is 
displayed in Fig.~\ref{fig:experiment}. Unfortunately, due to their 
large error bars, the present experimental data still allow for the $s$-wave 
approximation and the use of only BB terms in the elementary amplitude.
In the latter, it is obvious that experimental data at forward angles
with about 10\% error bars would be able to justify the important role of
the SB terms $F_6-F_{15}$.  
In the former, we note that the largest deviation of using
only $s$-waves does not appear at $W=4.10$ GeV. Instead, at the forward
directions the largest deviation (approximately 0.99 $\mu$b/sr,
see Fig.~\ref{fig:pwaves}) is found with $W=4.20$ GeV. Therefore, 
experimental data with the same quality, but
at $W=4.20$ GeV, would be very useful to check the validity of
the $s$-waves approximation.
For the sake of numerical accuracy and efficient cpu-time we would, however,
recommend the use of partial waves with $\alpha\le 4$ along with a full 
elementary amplitude.

The discrepancy between the calculated cross sections and 
the experimental data point at $\theta = 18.9^\circ$ requires 
a special explanation. At this kinematics we note that the
calculated cross sections are much smaller than those at 
the forward direction. This behavior seems to be almost 
independent of the total c.m. energy. However, we have also found
that the longitudinal part of the cross section 
($d\sigma_{\rm L}/d\Omega$) dominates other parts 
[see Eq.~(\ref{eq:ds_domega_virtual})]
in the whole kinematics shown in Fig.~\ref{fig:pwaves} and
falls off quickly as a function of the  $\theta$. On the
other hand, the angular distribution of the 
transverse cross section ($d\sigma_{\rm T}/d\Omega$) tends
to be more flat than the longitudinal one. As a consequence,
we may conclude that 
such behavior should originate from the elementary amplitude
and not from the effect of the nuclear wave functions. Thus,
if we believed that this experimental data point were 
correct, then we had to reconsider the improvement of
the elementary operator. As stated in the Introduction,
this means that the extraction of the elementary information from
the hypertriton production cross section would be mandatory in the
future works. Otherwise, new measurements at this kinematics
are urgently required.

\section{Conclusion}
\label{sec:conclusion}
In conclusion, we have investigated the effects of higher partial waves 
and relativistic terms on the accuracy of the calculated differential cross sections of the 
hypertriton electroproduction. We have shown that an accurate
calculation, with a maximum deviation of less than 4\%, could still be 
obtained if we used the three lowest partial waves with isospin zero
(i.e., using $\alpha\le 4$, since the selection rule excludes
the $\alpha=1$ component). 
In this case, the cpu-time for calculating differential cross sections 
is reduced by a factor of about 60. The exclusion of certain elementary amplitudes
$F_i$ has a tiny impact on the cpu-time, but a big impact on the accuracy
of the calculation. In view of this, for future consideration we suggest
the use of partial waves with $\alpha\le 4$ with a full elementary amplitudes
$F_i$. Comparison of the results with the available experimental data supports
our finding. New measurement of the hypertriton electroproduction
with about 10\% error bars would be very useful to clarify the
validity of these approximations as well as the importance of the
relativistic terms.

\section{Acknowledgements}
This work has been partially supported by the University of Indonesia.


\end{document}